\documentclass[conference]{IEEEtran}
\IEEEoverridecommandlockouts
\usepackage{cite}
\usepackage{amsmath,amssymb,amsfonts}
\usepackage{algorithmic}
\usepackage{array}
\usepackage{graphicx}
\usepackage{textcomp}
\usepackage{xcolor}
\usepackage[hyphens]{url}

\def\BibTeX{{\rm B\kern-.05em{\sc i\kern-.025em b}\kern-.08em
    T\kern-.1667em\lower.7ex\hbox{E}\kern-.125emX}}
\begin{document}

\title{Rethinking the Cloudonomics of Efficient I/O for Data-Intensive Analytics Applications}

\author{\IEEEauthorblockN{Chunxu Tang\IEEEauthorrefmark{1}\textsuperscript{\textsection}, Yi Wang\IEEEauthorrefmark{1}\textsuperscript{\textsection}, Bin Fan\IEEEauthorrefmark{1}, Beinan Wang\IEEEauthorrefmark{1}, Shouwei Chen\IEEEauthorrefmark{1}, Ziyue Qiu\IEEEauthorrefmark{2}\IEEEauthorrefmark{3}, \\Chen Liang\IEEEauthorrefmark{2}, Jing Zhao\IEEEauthorrefmark{2}, Yu Zhu\IEEEauthorrefmark{1}, Mingmin Chen\IEEEauthorrefmark{2}, Zhongting Hu\IEEEauthorrefmark{2}}
\IEEEauthorblockA{
\IEEEauthorrefmark{1}\textit{Alluxio, Inc.}, Foster City, California, USA\\
\IEEEauthorrefmark{2}\textit{Uber, Inc.}, San Francisco, California, USA\\
\IEEEauthorrefmark{3}\textit{Carnegie Mellon University}, Pittsburgh, Pennsylvania, USA
\\
\{chunxu.tang,hope.wang,bin,beinan,shouwei,david\}@alluxio.com,
ziyueqiu@cs.cmu.edu,\\ \{chliang,jingzhao,mingmin,curt\}@uber.com}
}

\maketitle
\begingroup\renewcommand\thefootnote{\textsection}
\footnotetext{The authors contribute equally to this work.}
\endgroup

\begin{abstract}

This paper explores a prevailing trend in the industry: migrating data-intensive analytics applications from on-premises to cloud-native environments. We find that the unique cost models associated with cloud-based storage necessitate a more nuanced understanding of optimizing performance. Specifically, based on traces collected from Uber's Presto fleet in production, we argue that common I/O optimizations, such as table scan and filter, and broadcast join, may lead to unexpected costs when naively applied in the cloud. This is because traditional I/O optimizations mainly focus on improving throughput or latency in on-premises settings, without taking into account the monetary costs associated with storage API calls. In cloud environments, these costs can be significant, potentially involving billions of API calls per day just for Presto workloads at Uber scale. Presented as a case study, this paper serves as a starting point for further research to design efficient I/O strategies specifically tailored for data-intensive applications in cloud settings.

\end{abstract}

\begin{IEEEkeywords}
cloudonomics, I/O, cloud, data analytics
\end{IEEEkeywords}

\section{Introduction}
\label{sec:intro}

This paper investigates a prevailing industry trend: migrating data-intensive analytics applications from on-premises environments to cloud-native infrastructures \cite{partlycloudy,uber-cloud,wu2021migrate}. By analyzing traces from production Presto queries at Uber, we argue that common I/O optimizations in these analytics applications may associate unexpected costs (on the order of tens of multiple million dollars per year) due to storage API calls as a part of the significantly different cost models presented by cloud vendors.
This oversight can lead to unexpectedly high costs and compromised efficiency when these applications are migrated to a cloud-based environment without proper adjustments. For successful migration and operation in the cloud, it is imperative for system researchers and developers to understand the nuances of cloud storage services and to devise efficient I/O strategies specifically tailored for applications within these environments.

Data-intensive analytics applications are epitomized by analytics computation engines such as Apache Spark, Apache Hive, Presto \cite{zaharia2010spark,thusoo2009hive,sun2023presto}, and others. Historically, these applications have been architected and operated within on-premises environments, relying on storage systems like HDFS (Hadoop Distributed File System), where strong data locality is ensured by minimizing data transfer between storage and computation.
Moreover, the discrete nodes or processes within a distributed computation engine typically make file-level requests individually, such as reading a file or listing a directory. 
The absence of coordination is based on the assumption that the costs associated with I/O requests are primarily incurred in transient resource consumption -- network bandwidth, memory, or processing power -- rather than as a direct financial expense.
For more than a decade, these design principles have guided the development of computation engines, leading and driving I/O optimizations that effectively improve I/O throughput and reduce latency.

However, as these applications transition to cloud environments, both the assumptions about data locality and cost models are put to the test.
First, cloud storage services such as AWS S3, Azure Blob Storage, and Google Cloud Storage (GCS) \cite{amazon-s3, azure-blob, google-gcs} ascend to prominence as the principal data lake solutions within their respective cloud ecosystems. As a result, data is no longer confined to a service on the local network; instead, it is served from cloud infrastructure, rendering the concept of node-level data locality obsolete.
Second, these cloud storage services impose pricing structures fundamentally different from traditional on-premises storage solutions. Specifically, they charge for discrete API calls, which are not necessarily proportional to the amount of data transferred~\cite{amazon-s3-pricing, azure-blob-pricing, gcs-pricing}. As a result, costs can accumulate quickly if this nuanced distinction is overlooked.

By examining data-intensive analytics workloads focusing on I/Os through the cost lens in a cloud environment, this paper makes the following contributions:
\begin{itemize}
    \item In Section~\ref{sec:cost}, we shed light on the importance of evolving our understanding and strategies in line with the distinct dynamics of cloud storage and its impact on application design and performance.
    \item In Section~\ref{sec:optimizations}, we present the case study, based on real-world traces from Uber, to illustrate that widely used I/O optimization techniques, such as table scan and filter, and broadcast join, may be linked with the unanticipated cost implication in the enterprise-grade cloud migration.
    \item In Section~\ref{sec:discussion}, we provide a fresh perspective on system design within the field of cloud computing to assist stakeholders in navigating the rapidly evolving landscape of data-intensive applications.
\end{itemize}

\section{Cloud Cost Model} \label{sec:cost}

\begin{table*}[ht]
\center
\caption{Storage cost for Amazon S3, Google Cloud Storage, and Azure Blob Storage. \\Pricing applies to AWS US East Ohio region, GCP North America regions, Azure West US 2 region. As of August 28, 2023.}
\label{tab:cloud-cost}
\begin{tabular}{l|lc}
  \textbf{Cloud storage} & \textbf{Operation Type} & \textbf{Price per 1,000 requests} \\ 
  \cline{1-3} 
  {Amazon S3 - S3 Standard}
   &
  \begin{tabular}[c]{@{}l@{}}PUT, COPY, POST, LIST requests\end{tabular} 
   & 
  \textbf{\$0.005} \\ 
  \cline{2-3}
  & \begin{tabular}[c]{@{}l@{}}GET, SELECT, and all other requests \end{tabular} 
  & \textbf{\$0.0004} \\ \hline
  
{GCS - Standard Storage - XML API} &
  \begin{tabular}[c]{@{}l@{}}GET Service\\ GET Bucket \footnote{when listing objects in \\ a bucket}\\ 
  PUT, POST \end{tabular} 
  &
  \textbf{\$0.005} \\
  \cline{2-3}
   &
  \begin{tabular}[c]{@{}l@{}}GET Bucket \footnote{when retrieving bucket configuration or when listing ongoing multipart uploads} \\ GET Object\\ HEAD \end{tabular} 
  & 
  \textbf{\$0.0004}\\
  \hline
\begin{tabular}[c]{@{}l@{}}Azure Blob Storage - \\ Standard (GPv2) Storage\end{tabular} 
&
  Write operations &
\begin{tabular}{cc}
    Premium & \textbf{\$0.00228} \\
    Hot & \textbf{\$0.0065} \\
    Cool & \textbf{\$0.013} \\
    Archive & \textbf{\$0.013} \\
\end{tabular}
 \\ \cline{2-3} 
 &
  Read operations 
 &
\begin{tabular}{cc}
    Premium & \textbf{\$0.00019} \\
    Hot & \textbf{\$0.0005} \\
    Cool & \textbf{\$0.0013} \\
    Archive & \textbf{\$0.65} \\
\end{tabular} \\ \hline
\end{tabular}
\end{table*}

This section describes the pricing models employed by cloud providers for storage resources.
Compared to conventional on-premises storage, the optional cost of accessing data is significant and often unclear to users largely contributed to the \emph{unpredictable} nature when running conventional data-intensive computation engines on modern cloud storage.
 
Table~\ref{tab:cloud-cost} compares the API call pricing models for accessing data across different cloud storage services such as S3, GCS, and Azure Blob Storage. A commonality among these vendors is that the cost of storage API calls depends on the number of API calls made, rather than the size of the data being accessed.
In terms of API call costs, it may be less expensive to access an object of 3 KB than to access the first and last KB of the same object, even though the latter approach requires only 2 KB in total while the former requests downloading 3 KB.

In addition, different cloud vendors categorize operations differently and employ varying service tiers that also influence costs. For example, Amazon S3 charges based on the number of requests, dividing them into two categories: PUT, COPY, POST, and LIST requests; and GET, SELECT, and all other requests. In contrast, Azure Blob Storage charges based on write and read operations. The choice of the specific storage tier—be it premium, hot, cool, or archive—can also have a significant impact on the overall cost.

In the context of running data-intensive applications like Hadoop, Spark, or Presto in the cloud, here is the gap we identified: Cloud storage services introduce a unique cost model that is based on the number of storage API calls and the amount of data transferred. These API calls for data-intensive applications, typically include operations like reading, writing, and accessing metadata, incur charges each time they are invoked. 
These systems were historically designed and optimized to minimize the amount of data transferred, not necessarily to reduce or consolidate the number of storage API calls. Additionally, the conventional I/O optimization approaches in data analytical systems, such as Presto, focus on read operations, which dominate industrial data analytical workloads \cite{uber-presto,tang2021forecasting}. As we are discussing the challenges and impact of these optimization approaches on the monetary cost during cloud migration, we concentrate on the read operations in this paper.

Data-intensive applications frequently involve extensive I/O activities, such as data processing and transfer, without any predefined rate limiting. This absence of limitations on I/O operations can lead to escalating costs, which can affect the financial feasibility of running such applications in the cloud.

\section{Impact of I/O Optimizations on Industrial Analytical Workloads}
\label{sec:optimizations}

\subsection{Cloud I/O Cost Challenges on Industrial Analytics Traffic Patterns}
\label{sec:optimizations:challenges}

\begin{figure}[thb]
  \centering
  \includegraphics[width=0.55\linewidth]{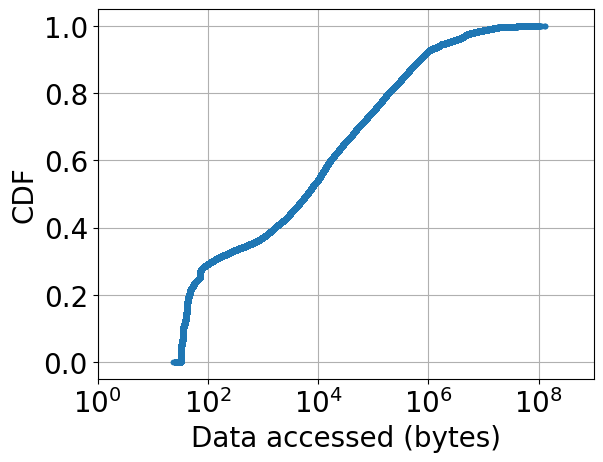}
  \caption{
  The cumulative distribution function (CDF) of data access sizes from storage on a typical Presto node at Uber.
  }
  \label{fig:uber-size-cdf}
\end{figure}

To understand the data access traffic patterns exhibited by analytics applications on large-scale, enterprise-level data and to analyze the impact of I/O optimizations on cloud costs after migration to a cloud-native environment, 
we collected traces from a production Presto cluster at Uber. Given the load balancing, the cluster's workload pattern is representative of Uber's offline Presto processing traffic, which is the main use case of Presto at Uber. Traces were amassed over roughly five days of cluster operations.
Combining I/O traces together with our operational experience in production, we make the following observations:

\begin{itemize}
    \item Analytics jobs must process an enormous volume of data read operations on a daily basis. Recent works from Meta \cite{sun2023presto}, Uber \cite{luo2022batch} and Twitter \cite{tang2021hybrid,tang2022serving} suggest that a Presto cluster can process tens to hundreds of petabytes of data per day.
    
    \item Column-oriented file formats dominate. Today, datasets for data analytics are primarily stored in column-oriented file formats, such as Apache Parquet and ORC \cite{vohra2016apache, apache-orc}. The primary reason for the popularity of column-oriented files over traditional row-based files is that they support optimizations like predicate pushdown, which improve data reading performance.

    \item Accessing small data fragments constitutes a major fraction of I/Os. At Uber, more than 50\% of Presto accesses on storage are less than 10 KB, and over 90\% are less than 1 MB, as shown in Figure \ref{fig:uber-size-cdf}. This is largely because datasets are stored in column-oriented file formats.   
\end{itemize}

Based on these observations, each read request in Presto workloads typically retrieves small data fragments, leading to billions of read API calls per day at Uber's scale and workload. Migrating the Presto workload to the cloud could, therefore, result in significant cost implications for analytics workloads. It is worth noting that this I/O traffic constitutes only a relatively small fraction of all data-intensive applications on a modern enterprise data platform. Understanding the sources and potential mitigation of these cost overheads due to API calls is critical.

Fundamentally, optimizations in widely-used analytics engines like Spark SQL and Presto are primarily geared towards improving query throughput and latency, driven by the need for fast and responsive data analysis. In on-premises settings, unlike in cloud-native environments, API calls don't have associated monetary costs. Specific optimization approaches, particularly those applied to column-oriented file formats, may reduce data reads but increase the number of data requests.

In the following section, we present a case study based on real-world data analytics workloads from Uber to analyze the impact of various I/O optimizations. It is worth noting that a range of popular optimizations, such as indexing, late materialization, and shuffle joins, are applied in Presto in production at Uber. However, due to page limitations, we will focus our case study on two typical optimization methods used in Uber's analytical workloads: table scans/filters in Section \ref{sec:optimizations:scan} and broadcast joins in Section \ref{sec:optimizations:join}. We aim to analyze their potential negative impact on billing costs in the absence of cost-oriented optimizations. Strategies to address these challenges will be discussed in Section \ref{sec:discussion}.

\subsection{Cost Challenges With Table Scan \& Filter Optimization}\label{sec:optimizations:scan}

\begin{figure}[htb]
  \centering
  \includegraphics[width=0.8\linewidth]{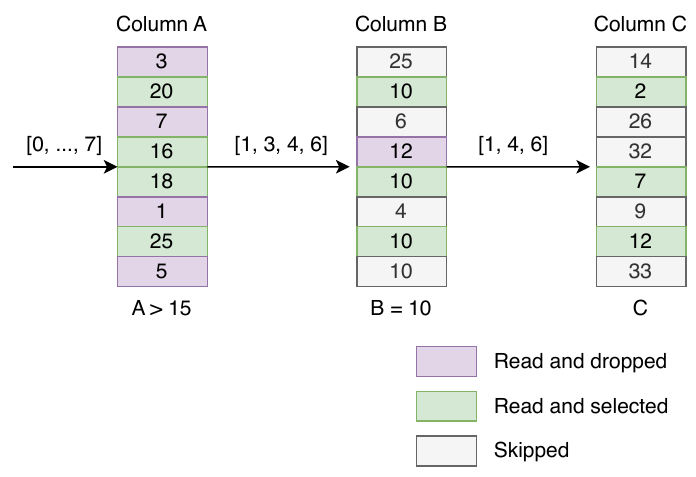}
  \caption{Applying filters to skip unnecessary rows in the processing of an SQL statement \textit{SELECT B, C FROM t WHERE $A > 15$ and B = 10}.}
  \label{fig:table-scan}
\end{figure}

One example of the focus on latency in optimizing analytical workload at Uber is the implementation and optimization of table scan operations. 
A table scan is a technique that reads all the rows of a table, which can be computationally expensive for large datasets.
To improve the performance of table scans, one major optimization is \emph{predicate pushdown}, commonly used in scanning column-oriented file formats, such as Apache Parquet and Apache ORC. This approach allows the database system to skip unnecessary rows and columns before retrieving the data from the disk. We demonstrate an example in Figure \ref{fig:table-scan}. Suppose the database system needs to process an SQL statement \textit{SELECT B, C FROM t WHERE $A > 15$ and B = 10}. After processing column A with condition \textit{$A > 15$}, the query engine passes the indexes 1, 3, 4, and 6 to the next stage as only the values of these indexes fulfill the requirement. Afterward, when processing column B, the query engine does not need to scan all rows, but only read the rows with the passed-down indexes. Other rows are skipped to avoid unnecessary scans. After checking with the condition \textit{B = 10}, the compute engine selects indexes 1, 4, and 6 and passes them to the next stage to select corresponding rows in column C. As a result, predicate pushdown can significantly reduce CPU and I/O usage, leading to faster query execution times and lower costs.

However, it's important to note that predicate pushdown can also have some drawbacks for cost when the table scan is hitting the cloud storage. While it saves CPU and I/O usage, it can create a larger number of smaller reads, which in turn can lead to more API calls to cloud storage. This can increase the overall cost of query execution, making it less cost-efficient in industrial use cases. At Uber, on average, each Presto data access fetches around 10 KB of data from underlying storage systems. As Uber's Presto system accesses around 10 PB to 50 PB of data daily, the Presto clusters would send 1,000 billion to 5,000 billion read requests to the storage per day.
By contrast, according to an analysis on Meta's Presto workloads \cite{sethi2019presto}, without the predicate pushdown optimization, the data read is around 5 times of the data read with the optimization. Applying that statistics to Uber's Presto workloads, without the predicate pushdown optimization, Uber's Presto system will accesses around 50 PB to 250 PB of data daily, and each Presto data access fetches 1 MB Parquet page (default Parquet setting at Uber). Consequently, without the predicate pushdown optimization, Uber's Presto will only send 50 billion to 250 billion read requests to the storage per day, only 5\% of the number of requests required when the optimization is enabled. To conclude, without additional cost-aware optimization approaches, the cloud migration will cause an unexpectedly higher API cost. Therefore, database systems need to carefully balance the benefits and costs of predicate pushdown when optimizing table scans for latency and cost efficiency.

\subsection{Cost Challenges With Broadcast Join Optimization}\label{sec:optimizations:join}

\begin{figure}[htb]
  \centering
  \includegraphics[width=0.9\linewidth]{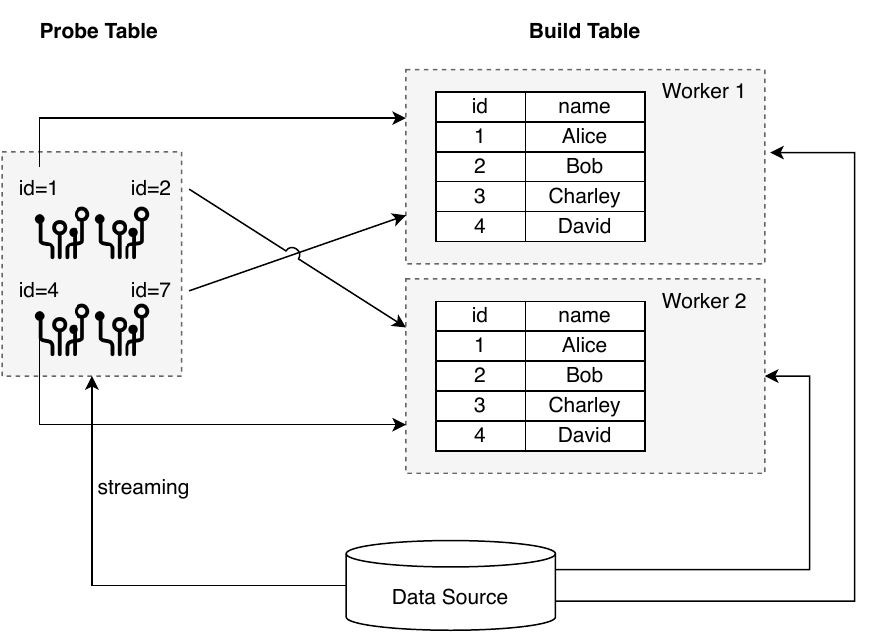}
  \caption{An example of broadcast join. Each worker loads the build table in memory and processes the items from the probe table in parallel.}
  \label{fig:broadcast-join}
\end{figure}

Another example of the focus on latency at Uber is the implementation and optimization of broadcast join operations in analytical workload. 
A broadcast join involves replicating a small table (build table) and sending it to all the nodes that have a part of the larger tables (probe table), which can significantly reduce the time and resources needed to complete the join. 
The concept is exemplified in Figure \ref{fig:broadcast-join}, where two compute worker nodes are involved in query processing. With the broadcast join enabled, both workers are responsible for loading the smaller build table into memory from the data source. Conversely, the larger probe table is loaded in a streaming pattern from the data source. The distributed processing pattern allows the two workers to handle the items in parallel, thereby expediting the overall processing speed.

With the focus on latency, there has been a shift in database systems towards optimizing broadcast joins for faster completion time. 
For example, Apache Spark can automatically optimize join operations and decide when to use broadcast joins based on the size of the tables and the available resources.

However, it's important to note that broadcast join may not always be the most cost-efficient approach, as each compute node needs to load all the data from the broadcasted table into memory or local disk storage. 
This can increase the storage cost of the system. 
Uber's Presto platform processes around 500 thousand queries per day, of which around 20\% of queries utilize the broadcast join optimization. In Presto, the default threshold for the build table size to trigger the optimization is 100 MB. From our observation, the build table sizes in Uber's analytical workloads are commonly large, ranging from 20 MB to 100 MB. At Uber, each Presto cluster contains 200 worker nodes, we can obtain 

\[200 \times 100MB \times 500k \times 20\% = 2 PB \]

This indicates that in one Uber's Presto cluster, constructing the build tables across workers only will need to access up to 2 PB of data. In addition, the size of data accessed is linear to the number of worker nodes in a Presto cluster, indicating that (1 - 1 / (number of nodes), aka 199 / 200 = 99.5\% of the 2 PB of data we calculated is wasted due to the duplicate broadcasted table reading efforts.
Considering that we are employing both table scan/filter and broadcast join optimization approaches, processing such an amount of data could result in more than 10 billion API calls per day without cost-oriented optimization.

\section{Discussion} \label{sec:discussion}

In this section, we discuss different strategies to address the challenges. 
\subsection{Improving Storage Resources Efficiency}

Improving the cost awareness and cost efficiency of the public cloud will be a winning recipe for both cloud vendors and their customers.
Currently, major cloud vendors offer a variety of storage services, providing the potential to utilize them in combination to achieve optimized performance and cost efficiency. For example, Amazon S3 Select \cite{s3-select} helps filter data in the storage layer, reduce the number of API calls, and decrease the size of data accessed, although it is restricted to AWS. Liu et al. \cite{liu2023cost} presented a comprehensive review of recent advancements in cost optimization strategies for cloud storage users, categorizing them into storage efficiency improvement, and utilization of cloud storage service features. 

The flexibility of these services allows for a tailored data storage strategy that meets individual business requirements, but also, when leveraged effectively, can result in significant cost savings. Park et al. \cite{park2020more} explored how to leverage burstable storage in the cloud to reduce costs and/or enhance performance for different I/O workloads having different data longevity requirements. Kadekodi et al. \cite{kadekodi2018case} pointed out that operating many small objects (kilobyte-sized) is expensive in cloud object stores due to the API call cost, and proposed a client-side change by combining multiple small objects into a single large object, at the potential cost of consistency. 
However, none of these works focus on data-intensive applications (such as analytics) or on the impact of widely used I/O optimizations on system designs.

\subsection{Creating a Virtual Storage Layer with Cache}

As an alternative strategy, the introduction of a platform as a coordinated virtual storage layer can help address these challenges. 1) It can eliminate redundant requests to the backend storage, reducing unnecessary data traffic and optimizing storage utilization. 2) It can incorporate rate-control mechanisms to prevent traffic throttling, thereby ensuring smooth and uninterrupted data operations. Moreover, we observed that in industrial data analytical workloads, hot data can be repeatedly accessed for hundreds of thousands of times within a single day. In Uber's data analytical workloads, 50\% of accessed data is accessed again in less than 2 hours; the top 10,000 data blocks attract around 90\% or more read traffic \cite{uber-hdfs}. 3) The virtual storage can preload hot data and serve such frequently accessed data with a cache service. 

There are some pioneering projects in this direction. For example, Wang et al. \cite{wang2022metadata} presented a unified metadata caching layer built on top of Presto, which caches intermediate results in data processing. For another instance, Li et al. \cite{li2018alluxio} proposed a Virtual Distributed File System (VDFS) called Alluxio to address the increasing complexities and challenges in data storage and processing, offering benefits like global data accessibility, efficient in-memory data sharing, high I/O performance, and flexible compute and storage choices.

However, cost efficiency is largely overlooked and data strategies are not well understood for this layer. Additionally, cache may cause a significant drawback: \emph{read amplification}, as compute engines often read only a small portion of data, but the granularity of a cache is often at the level of a file or block. We anticipate more studies in this direction to better understand the potential and limitations.

\subsection{Redesigning the Data-Intensive Applications to Be Cost-Aware}
Another effective strategy involves redesigning data-intensive applications focusing on I/O optimization, particularly through cost-based strategies. This involves fundamentally reassessing how applications interact with data and storage resources. Instead of relying on traditional methods of data access, it may be beneficial to architect these applications to exploit the strengths and work around the limitations of the chosen storage resources.

For example, Tan et al. \cite{tan2019choosing} did a pioneering study on choosing a cloud DBMS with an evaluation of both performance and cost, although they didn't explore in detail of the storage API cost discussed in this paper. Armbrust et al \cite{armbrust2021lakehouse} proposed Lakehouse architecture and argued that the industry is moving toward this architecture as a replacement for data warehouse.

Delta Lake, Iceberg, and Hudi \cite{armbrust2020delta,iceberg,hudi} introduce an ACID table storage layer and support row-level append and insert operations. However, this design could potentially cause a high number of small files and small read operations, which, as a consequence, could have an impact on the monetary cost of the cloud migration. We anticipate more research efforts invested in this direction to thoroughly evaluate the impact on cost with the latest data lake table layouts.
 
Overall, these system evolution and newly emerging datalake formats primarily focus on improving performance in cloud-native environments or providing stronger data consistency, while not necessarily considering I/O cost in the cloud as a major design driver. We believe it's critical to strike a careful balance between cost and performance, as this can lead to substantial improvements when implemented effectively.

\section{Conclusion}\label{sec:conclusion}

In conclusion, migrating data-intensive analytics applications to cloud infrastructures demands a reevaluation of standard optimizations that primarily focus on throughput and latency. This is due to the distinct cost models in cloud environments. For successful migration and sustainability, system researchers and developers must understand cloud storage services' nuances and devise efficient I/O strategies.

\bibliographystyle{IEEEtran}

\end{document}